\documentclass{PoS}

\usepackage{amsmath}
\usepackage{cite}

\title{The influence of the optical star on the jets of high-mass microquasars}

\ShortTitle{Influence of the star on HMMQ jets}

\author{\speaker{E. Molina} and V. Bosch-Ramon\\
        Departament de F\'isica Qu\`antica i Astrof\'isica, Institut de Ci\`encies del Cosmos (ICCUB),
        Universitat de Barcelona (IEEC-UB), Mart\' i i Franqu\`es 1, 08028 Barcelona, Spain\\
        E-mail: \email{emolina@fqa.ub.edu}}

\abstract{Microquasars are binary systems consisting of a compact object and a star that present relativistic jets. When the companion is a high-mass star, significant interaction between the stellar wind and the jets is expected. In this work, an overview of the different effects that a strong stellar wind may have in microquasar jets is given. Both analytical estimates and numerical simulations performed in the last years are reviewed. The results of a model for the non-thermal emission of such jets at large scales are also shown and discussed. Observational studies of two high-mass microquasars, Cygnus X-1 and Cygnus X-3, are compared to the model predictions.}

\FullConference{Frontier Research in Astrophysics - III (FRAPWS2018)\\
		28 May - 2 June 2018\\
		Mondello (Palermo), Italy}

\begin{document}

%####################################################
\section{Introduction}\label{intro}

Microquasars are binary systems that consist of a compact object (CO), either a black hole or a neutron star, which accretes matter from a companion star. The companion may be a low- or a high-mass star.  In the former case, the systems are called low-mass microquasars (LMMQ); in the latter, high-mass microquasars (HMMQ). The action of the accretion process, magnetic fields and possibly the CO rotation leads to the formation of bipolar jets launched from the CO, in which particles are accelerated to very high energies. These jets are typically seen in radio, although they are also a site for gamma-ray emission. For a classical review on microquasars, we refer the reader to \cite{mirabel99}.

In HMMQ, the stellar wind can have a strong effect on the jet propagation, both because the jet has to propagate through a medium filled with wind material, and because it may be both significantly bent from its initial direction, and recollimated, by the lateral impact of the wind. At large scales, orbital motion could also influence the dynamics of the jet by producing a helical pattern on the latter. This effect has been studied in \cite{bosch16}. Previously, several works studied the interaction of HMMQ jets with stellar winds by means of both analytical and numerical methods (e.g. \cite{romero05,perucho08,araudo09,perucho10,perucho12,bosch13,yoon15,yoon16}).

In this paper we review some previous works and study how the stellar wind and orbital motion affect the non-thermal emission of HMMQ jets. First, the different effects resulting from jet-wind interactions are studied both at the binary system scale and further away. Then, we present the results of a newly developed model for the large scale non-thermal emission of HMMQ jets, affected by strong stellar winds and orbital motion. Finally, a discussion is provided, with reference to the observations performed so far and future possibilities.

%####################################################
\section{Jet-wind interaction}

In what follows an overview of the different effects that the stellar wind can have in HMMQ jets is given by taking into account analytical and numerical results obtained by previous works. Jets are usually assumed to have, at their launching site (i.e. their base), a constant, mildly relativistic velocity $v_{\rm j}$ along the $z$-axis, and conical shape with half-opening angle $\theta_{\rm j}$. The numerical simulations discussed here and carried out in the past are typically performed for an O-type companion star with an isotropic wind. Values for the wind velocity and mass-loss rate of the star are in the range $v_{\rm w} = 2-2.5\times10^8$~cm~s$^{-1}$ and \ $\dot{M}_{\rm w} = 10^{-5}-10^{-6}$~M$_\odot$~yr$^{-1}$, respectively. The orbital separation is $a = 2-3\times10^{12}$~cm, similar to that inferred from the period in Cygnus X-1 (see, e.g. \cite{pooley99}). Different jet powers are considered in the simulations.

%****************************************************
\subsection{Recollimation shocks}

As the jet propagates away from its launching site, it may reach a point were its lateral ram pressure, proportional to $z^{-2}$, will become smaller than the wind ram pressure exerted on the jet, which remains approximately constant for $z < a$ (e.g. \cite{yoon16}). At this point, the jet lateral expansion in the star direction will be stopped by the wind, and an (asymmetric) recollimation shock will develop. It may be the case, however, that a very powerful jet never has a lateral pressure lower than its surrounding medium, preventing the formation of such a shock. Analytical approximations show that the condition for a recollimation shock to form is the following \cite{bosch16,yoon16}:
	\begin{equation}\label{Prec}
	P_{\rm j} \lesssim 2.4\times10^{37} \ \frac{\gamma_{\rm j} (\gamma_{\rm j}-1)}{\beta_{\rm j}}
    				   \left( \frac{\dot{M}_{\rm w}}{10^{-6}~{\rm M_\odot~yr^{-1}}} \right)
    				   \left( \frac{v_{\rm w}}{2\times10^{8}~{\rm cm~s^{-1}}} \right)
                       \ {\rm erg~s^{-1}} \ ,
	\end{equation}
where $P_{\rm j}$ is the jet power, $\gamma_{\rm j}$ is its Lorentz factor and $\beta_{\rm j} = v_{\rm j}/c$.

This condition is in accordance with the results of 3D relativistic simulations (e.g. \cite{perucho10}), which predict the formation of recollimation shocks at the binary scale and the development of significant instabilities downstream. Weak jets, with $P_{\rm j} \lesssim 10^{35}$~erg~s$^{-1}$, are disrupted very close to the CO, within a few times $10^{11}$~cm, due to the action of the stellar wind (see \cite{perucho08}, for 2D simulations). For very powerful jets, with $P_{\rm j} \gtrsim 10^{38}$~erg~s$^{-1}$ no recollimation is predicted by non-relativistic simulations \cite{yoon16}, as expected from Eq.~\eqref{Prec}.

%****************************************************
\subsection{Jet bending}

Jet bending due to the impact of the wind may also be computed in an analytical way. To this end, a parameter is introduced corresponding to the ratio between the wind momentum rate intercepted by the jet, and the jet momentum rate:
	\begin{equation}\label{chi}
	\chi_j \approx 0.3 \frac{\gamma_{\rm j}-1}{\gamma_{\rm j} \beta_{\rm j}}
    			   \left( \frac{P_{\rm j}}{10^{37}~{\rm erg~s^{-1}}} \right)
    			   \left( \frac{\theta_{\rm j}}{0.1~{\rm rad}} \right)
                   \left( \frac{\dot{M}_{\rm w}}{10^{-6}~{\rm M_\odot~yr^{-1}}} \right)
    			   \left( \frac{v_{\rm w}}{2\times10^{8}~{\rm cm~s^{-1}}} \right) \ .
	\end{equation}
From this expression, one can obtain the bending angle of the jet with respect to the $z$-axis as (see appendix in \cite{bosch16} for the system of equations to be solved):
	\begin{equation}\label{phi}
	\phi \approx \frac{\pi^2 \chi_{\rm j}}{2\pi\chi_{\rm j} + 4\chi_{\rm j}^{1/2} + \pi^2} \ .
	\end{equation}

Significant jet bending is observed in simulations by \cite{yoon15,yoon16} for jet powers between $10^{35}-10^{38}$~erg~s$^{-1}$. In the lower end of this range, bending angles of up to $\phi = 65^\circ$ are obtained, with the jet being totally disrupted as already discussed in the previous subsection. On the other hand, for the highest values of the jet power, the bending is only of a few degrees.

%****************************************************
\subsection{Orbital effects}

If the jet is not disrupted within the binary scale, the combination of orbital motion and jet bending results in a force, combination of the Coriolis effect plus wind ram pressure, that renders the jet helical. If the bending is significant, i.e. $\phi > \theta_{\rm j}$, this effect becomes  prominent. Otherwise, for $\phi < \theta_{\rm j}$, the conical expansion of the jet should smooth out the helical pattern and shape the geometry of the whole structure. We note that in these inequalities, and below, we consider for simplicity the jet expansion rate equal before and after bending within the binary system. Otherwise, the half-opening angle after recollimation is the one to be compared with $\phi$. %AIXO NO ESTA CLAR, PERO HO DEIXO

A helical pattern would appear in both a ballistic and a non-ballistic jet, although the presence of wind material in the medium favours the second option. In order for the jet to follow a ballistic trajectory, its power should satisfy the following condition \cite{bosch16}:
	\begin{equation}\label{Pbal}
	P_{\rm j} \gtrsim 10^{38} \ \frac{\gamma_{\rm j}-1}{\gamma_{\rm j} \beta_{\rm j}}
    				  \left( \frac{\theta_{\rm j}}{0.1~{\rm rad}} \right)
       				  \left( \frac{\dot{M}_{\rm w}}{10^{-6}~{\rm M_\odot~yr^{-1}}} \right)
    			   	  \left( \frac{v_{\rm w}}{2\times10^{8}~{\rm cm~s^{-1}}} \right) ^{1/3}
                      \ {\rm erg~s^{-1}} \ ,
	\end{equation}
which is not the case for the situations that we will study hereafter.

At small scales, however, the effect of orbital motion on the jet trajectory is negligible. A determination of the distance travelled by a non-ballistic jet in the star-CO direction before being significantly deviated due to Coriolis forces, may give us an idea of the scale at which orbital motion becomes relevant for jet propagation. A rough estimate of this distance, obtained by equating the jet momentum flux, and the wind momentum flux in the jet flow frame, is the following \cite{bosch16}:
	\begin{equation}\label{dturn}
	d_{\rm turn} \approx 1.8\times10^{13} \ \chi_{\rm j}^{-1/2}
               			 \left( \frac{v_{\rm w}}{2\times10^8~{\rm cm~s^{-1}}} \right)
                         \left( \frac{\omega}{1.4\times10^{-5}~{\rm rad~s^{-1}}} \right) \ {\rm cm} \ ,
	\end{equation}
where $\omega$ is the orbital angular velocity normalized to a period of 5 days. This expression is valid as long as the initial $\theta_{\rm j}$ is not significantly changed by interaction with the wind, and $d_{\rm turn} \gg a$. The distance along the $z$-axis where the orbital motion starts to dominate is just $z_{\rm turn} = d_{\rm turn}/\tan\phi$.

%####################################################
\section{Non-thermal emission from a helical jet}

A semi-analytical model for the emission of non-thermal radiation, from radio to gamma-rays, of helical jets affected by the stellar wind was developed in \cite{molina18}. The jet (counter-jet) is studied from the starting point of the helical structure, defined by $d_{\rm turn}$ and $z_{\rm turn}$ $(-z_{\rm turn})$, up to where the radiation output becomes very small. The particle distribution along the jet is computed considering synchrotron, inverse Compton (IC) with stellar photons, and adiabatic energy losses, and assuming a particle injection at the helical jet onset proportional to $E^{-2}$. The available power for accelerating non-thermal particles is taken as 1\% of the total jet power. Both synchrotron and IC emission mechanisms are considered, as well as gamma-ray absorption by the stellar photons via electron-positron pair production (see, e.g. \cite{gould67}). Different magnetic fields and system orientations with respect to a distant observer are studied. The main parameters used in the model are listed in Table~\ref{tab:parameters}. From Eq.~\eqref{phi}, this set of values yields $\phi \approx 20^\circ$. Moreover, condition in Eq.~\eqref{Pbal} is not met, meaning that the jet propagates in a non-ballistic way. For more details about the model we refer the reader to \cite{molina18}.

\begin{table}[h]
 \begin{center}
	\begin{tabular}{l c c}
    \hline \hline
	\multicolumn{1}{c}{Parameter}		&					&	Value							\\
    \hline
		Star temperature				& $T_\star$			&	$4\times10^4$ K 				\\
        Star luminosity 				& $L_\star$ 		& 	$10^{39}$ 	erg~s$^{-1}$		\\
        Wind speed 						& $v_{\rm w}$ 		&	$2\times10^8$~cm~s$^{-1}$		\\
        Wind mass loss rate 			& $\dot{M}_{\rm w}$ &	$10^{-6}$ M$_\odot$~yr$^{-1}$	\\
    	Jet power 						& $P_{\rm j}$ 		&	$3\times10^{36}$ erg~s$^{-1}$	\\
        Initial jet Lorentz factor 		& $\gamma_{\rm j}$ 	&	2								\\
        Half-opening angle 				& $\theta_{\rm j}$ 	&	0.1 rad							\\
    	Orbital separation 				& $a$ 				&	$3\times10^{12}$ cm 			\\ 	
    	Period 							& $P$ 				&	5 days							\\
        Distance to the observer 		& $d$ 				&	3 kpc 							\\
        \hline
    \end{tabular}
    \label{tab:parameters}
    \caption{List of the main parameters used in the non-thermal emission model.}
 \end{center}
\end{table}

\begin{figure}
   \begin{minipage}{0.48\linewidth}
	\centering
    \includegraphics[width=\linewidth]{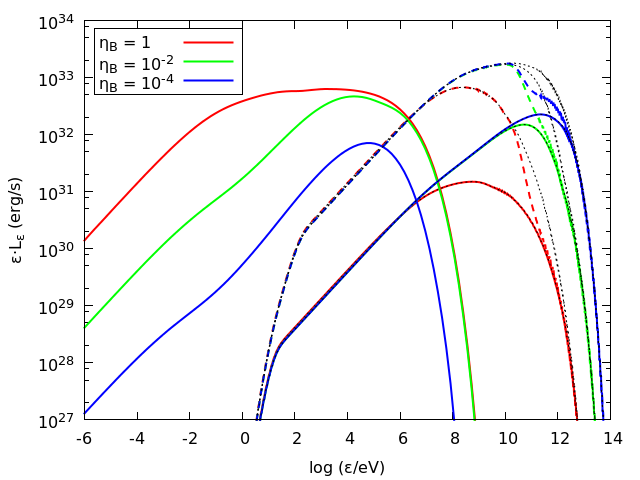}
    \caption{Synchrotron and IC SEDs of the jet (solid lines) and the counter-jet (dashed lines), for $i=30^\circ$, $\alpha=0.25$, and $\eta_{\rm B}=10^{-4}$ (blue lines), $10^{-2}$ (green lines), and 1 (red lines). Black dotted lines show the unabsorbed IC emission.}
    \label{fig:SEDB}
   \end{minipage}\hfill
   \begin{minipage}{0.48\linewidth}
    \centering
    \includegraphics[width=\linewidth]{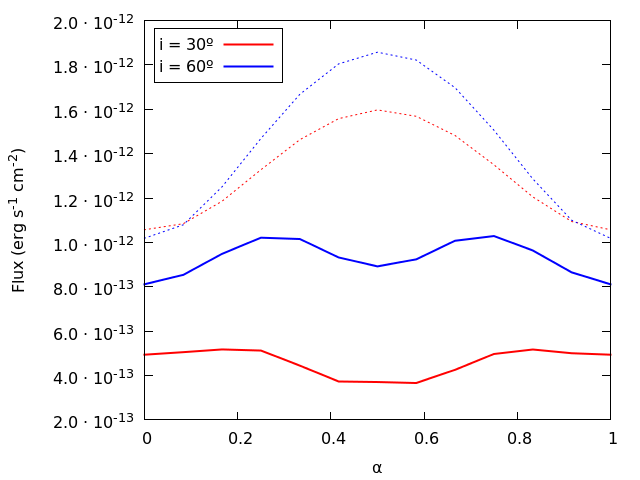}
    \caption{Light curve of total IC emission (jet + counter-jet) at $\varepsilon>100$~GeV, for $i=30^\circ$ (red lines) and $60^\circ$ (blue lines), and $\eta_B=10^{-2}$. The unabsorbed light curves are also shown (dotted lines).}
    \label{fig:Flux}
   \end{minipage}
\end{figure}

Figure~\ref{fig:SEDB} shows the synchrotron and IC spectral energy distributions (SED) of photons with energy $\varepsilon$ for different values of the magnetic field $B$ at the helical jet base. The jet and counter-jet SEDs are shown separately. The inclination of the system with respect to the observer is taken as $i=30^\circ$ and the orbital phase is $\alpha=0.25$, with $\alpha=0$ corresponding to the CO inferior conjunction (i.e. CO in front of the star), and $\alpha=0.5$ to the superior conjunction (CO behind the star). The effect of gamma-ray absorption, which is highly dependent on $i$ and $\alpha$, is also observed in the figure. The magnetic field at the helical jet base is parametrized through the ratio of magnetic pressure to stellar photon energy density, $\eta_B$, as:
	\begin{equation}\label{nB}
	\frac{B^2}{8\pi} = \eta_B \frac{L_\star}{4 \pi c D_\star^2} \ ,
	\end{equation}
where $D_\star$ is the distance from the star to the jet base.

The IC light curve for $\varepsilon>100$~GeV is shown in Fig. \ref{fig:Flux}. Both the absorbed and unabsorbed cases are shown for comparison, and two different inclinations are considered in order to assess the importance of the orientation in the IC emission and gamma-ray absorption. If absorption is not taken into account, emission peaks around the superior conjunction of the CO, where IC is more important. However, absorption via pair production with the stellar photons is also maximum around the same orbital phase, and for this reason a dip is obtained instead. The symmetric shape of the light curve is explained because most of the IC emission comes from the regions very close to the helical jet base. Nevertheless, if higher jet powers were considered, in accordance to Eqs.~\eqref{chi} and \eqref{dturn}, the helical structure would start further away from the binary system. In this case, IC emission would come from a more extended region, leading to an asymmetric light curve.

\begin{figure}
	\centering
	\includegraphics[width=\linewidth]{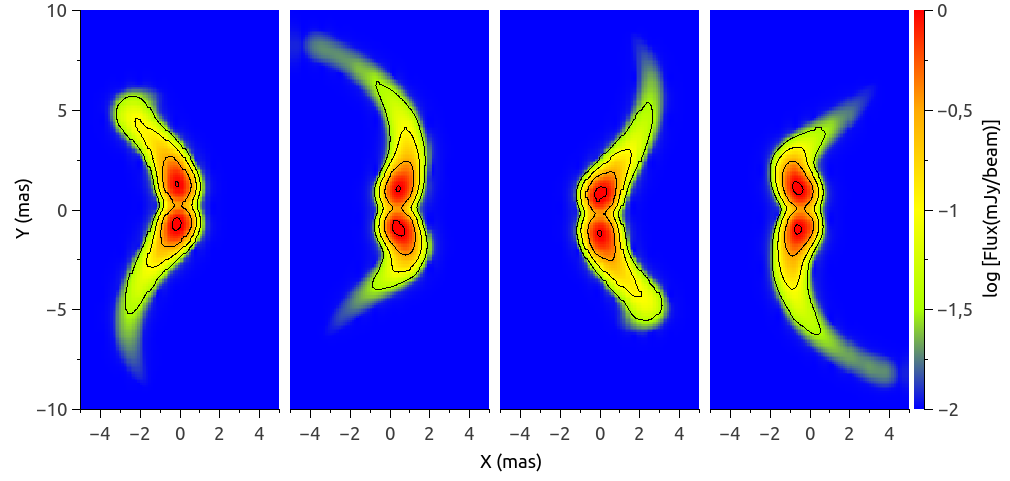}
	\caption{Sky maps at 5 GHz for $i$ = 60$^\circ$ and $\eta_B = 10^{-2}$. From left to right, orbital phases of 0, 0.25, 0.5 and 0.75 are shown. Contour levels: 0.032, 0.1, 0.32 and 1 mJy/beam. The beam size is 1~mas.}
	\label{fig:Sky60}
\end{figure}

Sky maps of the synchrotron emission at 5 GHz for $\eta_B=10^{-2}$, $i=60^\circ$, and 4 different orbital phases, are shown in Fig.~\ref{fig:Sky60}. In order to distribute the emission along the jet over its physical sky-projected area (our model gives just the path of the helical expanding jet), the emission is convolved with a 2D Gaussian with standard deviation $\sigma = r_{\rm j}/2$, where $r_{\rm j}$ is the jet radius at each point. Moreover, the resulting maps are again convolved with a Gaussian with FWHM = 1~mas in order to mimic the response of a radio, very long baseline interferometer (see, e.g. \cite{walker95}). The total received spectral flux density for each map is 7.4~mJy, within the sensitivity limits of current instrumentation. Note that this number would increase for a higher magnetic energy parameter as $\propto \eta_B^{3/4}$.

%####################################################
\section{Summary and discussion}

We have shown different ways in which a stellar wind could affect the jets in a HMMQ.  Analytical estimates and hydrodynamical 3D simulations predict the formation of recollimation shocks for moderate jet powers, which could be efficient sites for particle acceleration. The combination of jet bending by the lateral impact of the wind, and orbital motion, makes a non-ballistic helical jet a likely possibility in HMMQ, as already stated by \cite{bosch16}. There are currently 2 known sources that could fulfil the conditions for such a jet pattern to manifest: Cygnus X-1 and Cygnus X-3. Radio observations of the latter jets \cite{mioduszewski01,miller04} seem to indicate that such a behaviour is taking place, although the causes of the helical pattern are still uncertain. For Cygnus X-1, observations are not clear enough to determine whether a helical jet is present or not \cite{stirling01}.

Significant emission from radio to gamma-rays is predicted by the model for the non-thermal radiation of a helical jet, even for a modest non-thermal energy fraction (1\%). Most of this emission originates at the helical jet onset, where synchrotron and IC mechanisms radiate very efficiently. For a source at a distance of a few kpc, however, fluxes detectable with current instrumentation are predicted coming from an extended region, in which a hint of a helical structure may be appreciated in the light curve and the radio morphology. This is so even for moderate magnetic fields. Modulation of the gamma-ray emission due to orbital motion, with important absorption effects, is expected from HMMQ with a $d_{\rm turn}$-value similar to that considered in this work. Both the radio morphology and the gamma-ray light curve, even when this model does not try to fit any specific source, are not in contradiction with observations of Cygnus X-1 and Cygnus X-3 \cite{mioduszewski01,miller04,zanin16,zdziarski18}. More detailed modelling, along with future high-resolution radio observations, and high-energy light curves, could provide a way to properly characterize and disentangle the emission from a helical jet.

\bigskip
\bigskip
%####################################################
\acknowledgments
The authors want to thank the organizers for their kind invitation and for organizing the scientific workshop. We acknowledge support by the Spanish Ministerio de Econom\'{i}a y Competitividad (MINECO/FEDER, UE) under grant AYA2016-76012-C3-1-P, with partial support by the European Regional Development Fund (ERDF/FEDER), MDM-2014-0369 of ICCUB (Unidad de Excelencia `Mar\'{i}a de Maeztu'), and the Catalan DEC grant 2017 SGR 643. EM acknowledges support from MINECO through grant BES-2016-076342.

\bibliographystyle{JHEP}
\bibliography{references}

\providecommand{\href}[2]{#2}\begingroup\raggedright\begin{thebibliography}{10}

\bibitem{mirabel99}
I.~F. {Mirabel} and L.~F. {Rodr{\'{\i}}guez}, \emph{{Sources of Relativistic
  Jets in the Galaxy}},
  \href{https://doi.org/10.1146/annurev.astro.37.1.409}{\emph{ARA\&A}
  {\bfseries 37} (1999) 409}
  [\href{https://arxiv.org/abs/astro-ph/9902062}{{\ttfamily
  astro-ph/9902062}}].

\bibitem{bosch16}
V.~{Bosch-Ramon} and M.~V. {Barkov}, \emph{{The effects of the stellar wind and
  orbital motion on the jets of high-mass microquasars}},
  \href{https://doi.org/10.1051/0004-6361/201628564}{\emph{A\&A} {\bfseries
  590} (2016) A119} [\href{https://arxiv.org/abs/1604.06360}{{\ttfamily
  1604.06360}}].

\bibitem{romero05}
G.~E. {Romero} and M.~{Orellana}, \emph{{Gamma-ray and neutrino emission from
  misaligned microquasars}},
  \href{https://doi.org/10.1051/0004-6361:20052664}{\emph{A\&A} {\bfseries 439}
  (2005) 237} [\href{https://arxiv.org/abs/astro-ph/0505287}{{\ttfamily
  astro-ph/0505287}}].

\bibitem{perucho08}
M.~{Perucho} and V.~{Bosch-Ramon}, \emph{{On the interaction of microquasar
  jets with stellar winds}},
  \href{https://doi.org/10.1051/0004-6361:20078929}{\emph{A\&A} {\bfseries 482}
  (2008) 917} [\href{https://arxiv.org/abs/0802.1134}{{\ttfamily 0802.1134}}].

\bibitem{araudo09}
A.~T. {Araudo}, V.~{Bosch-Ramon} and G.~E. {Romero}, \emph{{High-energy
  emission from jet-clump interactions in microquasars}},
  \href{https://doi.org/10.1051/0004-6361/200811519}{\emph{A\&A} {\bfseries
  503} (2009) 673} [\href{https://arxiv.org/abs/0906.4803}{{\ttfamily
  0906.4803}}].

\bibitem{perucho10}
M.~{Perucho}, V.~{Bosch-Ramon} and D.~{Khangulyan}, \emph{{3D simulations of
  wind-jet interaction in massive X-ray binaries}},
  \href{https://doi.org/10.1051/0004-6361/201014241}{\emph{A\&A} {\bfseries
  512} (2010) L4} [\href{https://arxiv.org/abs/1002.4562}{{\ttfamily
  1002.4562}}].

\bibitem{perucho12}
M.~{Perucho} and V.~{Bosch-Ramon}, \emph{{3D simulations of microquasar jets in
  clumpy stellar winds}},
  \href{https://doi.org/10.1051/0004-6361/201118262}{\emph{A\&A} {\bfseries
  539} (2012) A57} [\href{https://arxiv.org/abs/1112.2520}{{\ttfamily
  1112.2520}}].

\bibitem{bosch13}
V.~{Bosch-Ramon}, \emph{{Relativistic stellar jets: dynamics and non-thermal
  radiation. Estimates on the dynamical impact of the stellar wind in low- and
  high-mass microquasars}},  in \emph{European Physical Journal Web of
  Conferences}, vol.~61, p.~03001, 2013,
  \href{https://arxiv.org/abs/1312.2377}{{\ttfamily 1312.2377}},
  \href{https://doi.org/10.1051/epjconf/20136103001}{DOI}.

\bibitem{yoon15}
D.~{Yoon} and S.~{Heinz}, \emph{{Global Simulations of the Interaction of
  Microquasar Jets with a Stellar Wind in High-mass X-ray Binaries}},
  \href{https://doi.org/10.1088/0004-637X/801/1/55}{\emph{ApJ} {\bfseries 801}
  (2015) 55} [\href{https://arxiv.org/abs/1501.03827}{{\ttfamily 1501.03827}}].

\bibitem{yoon16}
D.~{Yoon}, A.~A. {Zdziarski} and S.~{Heinz}, \emph{{Formation of recollimation
  shocks in jets of high-mass X-ray binaries}},
  \href{https://doi.org/10.1093/mnras/stv2954}{\emph{MNRAS} {\bfseries 456}
  (2016) 3638} [\href{https://arxiv.org/abs/1508.04539}{{\ttfamily
  1508.04539}}].

\bibitem{pooley99}
G.~G. {Pooley}, R.~P. {Fender} and C.~{Brocksopp}, \emph{{Orbital modulation
  and longer term variability in the radio emission from Cygnus X-1}},
  \href{https://doi.org/10.1046/j.1365-8711.1999.02225.x}{\emph{MNRAS}
  {\bfseries 302} (1999) L1}
  [\href{https://arxiv.org/abs/astro-ph/9809305}{{\ttfamily
  astro-ph/9809305}}].

\bibitem{molina18}
E.~{Molina} and V.~{Bosch-Ramon}, \emph{{Non-thermal emission from high-mass
  microquasar jets affected by orbital motion}}, {\emph{submitted for
  publication in A\&A} (2018) }.

\bibitem{gould67}
R.~J. {Gould} and G.~P. {Schr{\'e}der}, \emph{{Opacity of the Universe to
  High-Energy Photons}},
  \href{https://doi.org/10.1103/PhysRev.155.1408}{\emph{Physical Review}
  {\bfseries 155} (1967) 1408}.

\bibitem{walker95}
R.~C. {Walker}, \emph{{What the VLBA Can Do For You}},  in \emph{Very Long
  Baseline Interferometry and the VLBA} (J.~A. {Zensus}, P.~J. {Diamond} and
  P.~J. {Napier}, eds.), vol.~82 of \emph{Astronomical Society of the Pacific
  Conference Series}, p.~133, 1995.

\bibitem{mioduszewski01}
A.~J. {Mioduszewski}, M.~P. {Rupen}, R.~M. {Hjellming}, G.~G. {Pooley} and
  E.~B. {Waltman}, \emph{{A One-sided Highly Relativistic Jet from Cygnus
  X-3}}, \href{https://doi.org/10.1086/320965}{\emph{ApJ} {\bfseries 553}
  (2001) 766} [\href{https://arxiv.org/abs/astro-ph/0102018}{{\ttfamily
  astro-ph/0102018}}].

\bibitem{miller04}
J.~C.~A. {Miller-Jones}, K.~M. {Blundell}, M.~P. {Rupen}, A.~J. {Mioduszewski},
  P.~{Duffy} and A.~J. {Beasley}, \emph{{Time-sequenced Multi-Radio Frequency
  Observations of Cygnus X-3 in Flare}},
  \href{https://doi.org/10.1086/379706}{\emph{ApJ} {\bfseries 600} (2004) 368}
  [\href{https://arxiv.org/abs/astro-ph/0311277}{{\ttfamily
  astro-ph/0311277}}].

\bibitem{stirling01}
A.~M. {Stirling}, R.~E. {Spencer}, C.~J. {de la Force}, M.~A. {Garrett}, R.~P.
  {Fender} and R.~N. {Ogley}, \emph{{A relativistic jet from Cygnus X-1 in the
  low/hard X-ray state}},
  \href{https://doi.org/10.1046/j.1365-8711.2001.04821.x}{\emph{MNRAS}
  {\bfseries 327} (2001) 1273}
  [\href{https://arxiv.org/abs/astro-ph/0107192}{{\ttfamily
  astro-ph/0107192}}].

\bibitem{zanin16}
R.~{Zanin}, A.~{Fern{\'a}ndez-Barral}, E.~{de O{\~n}a Wilhelmi},
  F.~{Aharonian}, O.~{Blanch}, V.~{Bosch-Ramon} et~al., \emph{{Gamma rays
  detected from Cygnus X-1 with likely jet origin}},
  \href{https://doi.org/10.1051/0004-6361/201628917}{\emph{A\&A} {\bfseries
  596} (2016) A55} [\href{https://arxiv.org/abs/1605.05914}{{\ttfamily
  1605.05914}}].

\bibitem{zdziarski18}
A.~A. {Zdziarski}, D.~{Malyshev}, G.~{Dubus}, G.~G. {Pooley}, T.~{Johnson},
  A.~{Frankowski} et~al., \emph{{A comprehensive study of high-energy gamma-ray
  and radio emission from Cyg X-3}},
  \href{https://doi.org/10.1093/mnras/sty1618}{\emph{MNRAS} {\bfseries 479}
  (2018) 4399} [\href{https://arxiv.org/abs/1804.07460}{{\ttfamily
  1804.07460}}].

\end{thebibliography}\endgroup

\bigskip
\bigskip
\noindent {\bf DISCUSSION}

\bigskip
\noindent {\bf J. BEALL'S QUESTION:} When you show the helical structure of the jet, is that just density or emissivity?

\bigskip
\noindent {\bf E. MOLINA'S ANSWER:} The sky map where the helical structure is seen shows the synchrotron flux at 5~GHz. This flux is obtained after convolving the jet emission with a Gaussian beam with FWHM = 1 mas.

\end{document}